\documentclass{article}
\usepackage{amsmath,amssymb,upgreek,appendix,enumerate,graphicx}

\newcommand{\re}{\mathrm{Re}\;}
\newcommand{\im}{\mathrm{Im}\;}
\newcommand{\II}{\mathbf{I}}

\newcommand{\D}{\mathrm{D}}
\newcommand{\dd}{\mathrm{d}}
\newcommand{\e}{\mathrm{e}}
\newcommand{\ii}{\mathrm{i}}
\renewcommand{\u}{\mathbf{u}}
\newcommand{\x}{\mathbf{x}}
\newcommand{\Div}{\mathrm{div\;}}
\newcommand{\sgn}{\mathrm{sgn\;}}

\newtheorem{definition}{Definition}
\newtheorem{theorem}{Theorem}

\date{}

\begin{document}

\author{A. HANYGA\\ ul. Bitwy Warszawskiej 14 m. 52, 02-366 Warszawa, Poland\\
{\tt ajhbergen@yahoo.com}}

\title{DISPERSION AND ATTENUATION FOR AN
ACOUSTIC WAVE EQUATION CONSISTENT WITH VISCOELASTICITY}

\maketitle

\begin{abstract}    
An acoustic wave equation for pressure accounting for viscoelastic attenuation is derived
from viscoelastic equations of motion. It is assumed that the relaxation moduli are completely monotonic.
The acoustic equation  differs significantly from the equations proposed by Szabo (1994) and in several 
other papers. Integral representations of dispersion and attenuation are derived. General 
properties and asymptotic behavior of attenuation and dispersion in the low and high frequency range 
are studied. The results are compatible with experiments. The relation between the asymptotic
properties of attenuation and wavefront singularities is 
examined.
The theory is applied to some classes of viscoelastic models 
and to the quasi-linear attenuation reported in seismology.
\end{abstract}
\textbf{keywords:} wave propagation; viscoelasticity; attenuation; completely monotonic 
functions; complete Bernstein functions; ultrasound

\section*{Notation.}

{\small
\begin{tabular}{lll}
$f\ast g $ & convolution & $\int_0^t f(s)\, g(t-s) \, \dd s$\\
$\mathcal{L}(f) = \tilde{f}$ & Laplace transform of $f$ & $\int_0^\infty \e^{-p t}\, f(t)\, \dd t$\\
$]a,b]$ & the set $a < x \leq b$ & \\
$f(x) \sim_a g(x)$ & asymptotic equivalence & $0 < \lim_{x\rightarrow a} f(x)/g(x) < \infty $ for $a = 0$ or $\infty$.
\end{tabular}
}

\section{Introduction.}
 
Correct modeling 
of wave attenuation is an important objective for several branches of acoustics and a vast 
literature is dedicated to this subject. We shall be concerned here with 
modeling of intrinsic attenuation (i.e. excluding attenuation due to backscatter) in the context of 
pressure wave equations used in acoustics.
In this context many attenuation models have 
been constructed ad hoc to match some aspect of the experimental data without taking 
into account the physical nature of the wave motion involved.

In many applications the mathematical model of propagation of acoustic pulses (longitudinal waves) 
in bio-tissues and polymers is based on  
a linear wave equation for the pressure field. The pressure wave equation can
be derived from the equations of motion of linear elasticity. In order to account
for attenuation some authors included in the wave equation an additional term involving a fractional time derivative or a more 
general pseudo-differential operator acting on the time variable \cite{Szabo1,Szabo2,SzaboWu00}. This 
method of accounting for attenuation is called the time causal method in \cite{Szabo}.
In this context causality 
means that the pseudo-differential operator is the convolution with a causal
distribution, which ensures the validity of Kramers-Kronig relations. In \cite{KellyMcGoughMeerschaert08,StrakaMeerschaertMcGough2013} the time derivatives have been modified in order to ensure that the attenuation obeys a fixed power law with 
an exponent $> 1$ in the entire frequency range. 
It will be shown that in all these cases the  
resulting pressure wave equations are inconsistent with the viscoelastic equations 
of motion. 

Introducing attenuation by modifying the time derivatives in the equations of motion is justified in the context of poroelasticity \cite{Biot:VE,HanICTCAThm:BIOT,LokshinRokDAN,HanRok,FastTrack,LuHanygaGP,LuHanygaPorous1}.
In poroelasticity and poroacoustics the attenuation due to viscous flow in a porous medium is represented by 
the viscodynamic operator, which is a pseudo-differential operator acting on the 
time variables \cite{Norris1:BIOT,PrideGangiDaleMartin,PrideBerrymanHarris}. 
In poroelastic equations of motion the viscodynamic operator is applied to the inertial terms. 
In viscoelasticity attenuation is introduced through a time convolution operator 
in the constitutive equations. It will be shown in
Sec.~\ref{sec:derivation} that in both the viscoelastic equations of motion and in the 
acoustic wave equation attenuation  
is represented by a term involving a pseudodifferential operator acting on both the 
time and spatial variables. This observation is not new, for example Stokes' equation 
has this structure.

The attempts to explain
the experimentally observed power law frequency dependence of attenuation in polymers, 
bio-tissues and some viscous fluids in terms of an oversimplified power law attenuation 
model \cite{Szabo1,KellyMcGoughMeerschaert08,StrakaMeerschaertMcGough2013}
result in unbounded phase speeds. Experiments covering the range 1-- 250\, MHz
indicate that the exponent of the 
power law lies 
between 1 and 2. In the oversimplified model it is assumed that the power law extends to the 
high frequency
range. A power law with an exponent $> 1$ in the high-frequency range entails wave propagation with 
unbounded phase speed. It has been shown in \cite{HanWM2013} that experimental
data lie in the low frequency range represented by the low-frequency asymptotics of 
viscoelastic relaxation models. It is shown in \cite{HanWM2013}  
that viscoelastic solids (i.e. viscoelastic media in which strain under constant load does not
relax to 0) have power law exponents in the range 1--2 in the low-frequency range. 
Exponents below one indicate that the material is a viscoelastic fluid - such as
some bio-tissues subject to a constant shear load
These general results are is corroborated by the analysis of specific relaxation models in 
\cite{HolmSinkus2010} for longitudinal waves. This indicates that experimental
data can be explained in terms of viscoelastic models.
 
Kowar,
Schertzer and coworkers \cite{KowarScherzerBonnefond2011,KowarSchertzer2012} 
constructed lossy wave equations with  
a bounded propagation speed by applying causal attenuation operators to the
elastic pressure wave equation, which ensures bounded 
phase speed. They found that superlinear attenuation rates exist only 
in the low frequency range covered by the experiments. In their model,  
attenuation exhibits sublinear frequency dependence in the high frequency range. 
This approach however amounts to a modification of the time derivatives in
the pressure wave equation, which is
not consistent with viscoelastic equations of motion.

If the pressure wave equation is derived from viscoelastic equations of motion but the attenuation 
grows at a superlinear rate in the high frequency range then the creep compliance is not 
concave \cite{HanJMP2013}. This contradicted by observations.
In the special case of strict power law attenuation with an exponent $> 1$ in the
entire frequency range it is easy to 
prove that the creep compliance is decreasing and convex. 
 
Several researchers have avoided the problem by working 
directly with the equations 
of motion of linear viscoelasticity or thermo-viscoelasticity with an appropriate 
stress relaxation model \cite{NachmanAl90,EndeLionLammering2011,NasholmHolm2011}. 
We shall however present an approximate derivation of a linear pressure wave equation 
based on the equations of motion of 
viscoelasticity. It will be thus demonstrated that viscoelastic attenuation is 
represented in the pressure wave equation by a {\em mixed} temporal and spatial derivative 
or a time convolution operator acting on a Laplacian of the pressure field 
(Sec.~\ref{sec:appl}).
It will then be shown in Sec.~\ref{sec:disp-att} that the attenuation and dispersion associated with
the wave equation has the same properties as the attenuation and dispersion
in linear viscoelasticity examined in \cite{SerHanJMP,HanWM2013}. The theory will
be applied to some classes of viscoelastic media.

In Sec.~\ref{sec:regularity} we shall establish the relations between the asymptotics 
of the attenuation function and regularity properties of Green's function at the wavefronts. So far regularity at the wavefronts has only been 
studied in connection with the singularity of the relaxation modulus or its derivative.

The dispersion-attenuation theory will also be used to examine the controversial 
linear frequency dependence of attenuation observed in 
geological media \cite{Futterman,Buckingham1,Jiang2010}. 
Linear attenuation is incompatible with viscoelasticity and with bounded
phase speed. 
We shall therefore look for the closest approximation to linear attenuation compatible 
with bounded phase speed. We investigate here attenuation models which differ
from linear attenuation by a slowly varying factor.

In Sec.~\ref{sec:derivation} a wave equation for pressure in a viscoelastic medium 
will be derived. In Sec.~\ref{sec:appl} the dispersion and attenuation theory will 
be developed for this equation.
The dissipation-attenuation theory presented in Sec.~\ref{sec:disp-att} depends 
on the assumption that the viscoelastic relaxation modulus is
a completely monotonic (CM) function \cite{HanDuality}. The CM property is so universal in
viscoelasticity that several attempts have been made to justify it
by an {\em a priori} argument \cite{BerisEdwards93,Day70b,AnderssenLoy02a}. In particular every
spring-dashpot model and every fractional generalization of such models has
a completely monotonic relaxation modulus. The last statement is an easy consequence 
of the duality theorem proved in \cite{HanDuality}. Many other models of 
viscoelastic relaxation (Prony series, the Havriliak-Negami model and its special 
cases) are also expressed in terms CM relaxation moduli. 

In Sec.~\ref{sec:regularity} it is shown that the order of the wavefront singularity 
in media with bounded wavefront speed 
depends on the asymptotic behavior of the attenuation function at infinity.

In Sec.~\ref{sec:examples} the general theory is applied to some examples 
involving strongly singular convolution operators
($K$ is singular at 0 but locally integrable, the phase speed is unbounded, Sec.~\ref{sec:stronglysingular}), 
weakly singular kernels 
($K$ is non-singular, but its derivative $K^\prime$ has a singularity at 0, Sec.~\ref{sec:powerlaw}) and 
non-singular 
kernels ($K$ and $K^\prime$ are continuous at 0, the attenuation function is bounded). In the first case 
the phase speed is unbounded while in the second case the wave fields exhibit 
finite wavefront speed. In Section~\ref{sec:ql} the case of nearly linear 
frequency dependence of attenuation is examined.

\section{Derivation of the pressure wave equation.}
\label{sec:derivation}

Consider a homogeneous isotropic compressible viscous fluid defined by the constitutive 
equation 
\begin{gather}  \label{eq:vfconst}
\upsigma = - P \, \II + \lambda \, \Div \u_{,t}\,\II + \mu \,\left[ \nabla \u_{,t} + \left( \nabla \u_{,t}\right)^\mathsf{T} \right]\\
P = c_0^{\;2} \, (\rho - \rho_0) \label{eq:vfstate}
\end{gather}
where $\upsigma$ denotes the Cauchy stress tensor, $P$ represents the elastic part 
of the pressure, $\u$ is the displacement vector, $\rho$ is the density and $\rho_0$ 
is a reference density. The parameters $K := \lambda + 2 \mu$ and $\mu$ are dynamic viscosities
of volumetric and shear deformations and $c_0$ is an elastic propagation speed.

Mass conservation can be expressed in the form  
$\tau_{,t} = \tau \, \Div \u_{,t}$, where $\tau := 1/\rho$. Hence 
\begin{equation}  \label{eq:vfmass}
\rho_{,t} = - \rho \, \Div \u_{,t}
\end{equation}
For further reference we note that $\rho - \rho_0 = \rho_0\, [\exp(-\Div \u) - 1]  \approx 
- \rho_0\, \Div \u$ and thus equation~\eqref{eq:vfstate} assumes the
form $P \approx -\mathcal{K}\, \Div \u$,
$\mathcal{K} = \rho_0\, c_0^{\;2}$,  in the linear approximation.

Equation~\eqref{eq:vfmass} implies that
 $$\Div \upsigma = -\nabla P + (\lambda + \mu)\, \nabla \, \Div \u_{,t} + \mu \nabla^2 \, \u_{,t}$$
The momentum balance $\rho \, \u_{,tt} = \Div \upsigma $ and the mass balance 
imply the following equation for $\xi := \Div \u$:
\begin{equation} \label{eq:ytrx}
\rho \, \xi_{,tt} = - \nabla^2\,P  + (\lambda + 2 \mu) \, \nabla^2\, \xi_{,t}
\end{equation}
Eqs~\eqref{eq:vfstate} and \eqref{eq:vfmass} imply that $c_0^{\;-2} \, P_{,t} = -\rho\, \xi_{,t}$ and
$c_0^{\;-2} \, P_{,tt} = -\rho\,\xi_{,tt} - \rho_{,t} \, \xi_{,t} = -\rho\, \left( \xi_{,tt} - \xi_{,t}^{\;2}\right)$. 
In a linearized theory we assume that
\begin{equation}
\vert \xi_{,t}\vert^2 \ll \vert \xi_{,tt} \vert
\end{equation}
hence $c_0^{\;-2} \, P_{,tt} = -\rho\,\xi_{,tt}$. The final pressure equation is 
obtained by substituting the last results in \eqref{eq:ytrx}
\begin{equation} \label{eq:pressure}
c_0^{\;-2}\, P_{,tt} = \nabla^2 \, P + \frac{\lambda + 2 \mu}{\rho_0\, c_0^2} \nabla^2\,P_{,t}
\end{equation}

If multiplication by $\lambda$ and $\mu$ in eq.~\eqref{eq:vfconst} is replaced by convolutions
with CM kernels $\lambda(t)$ and $\mu(t)$, respectively, then eq.~\eqref{eq:pressure} assumes
the following form
\begin{equation} \label{eq:pressure-attenuation}
c_0^{\;-2} P_{,tt} = \nabla^2 \, P + (\rho_0\, c_0^2)^{-1}  \nabla^2\,K\ast P_{,t}
\end{equation} 
for $x \in \mathbb{R}^d$,$d =1$ or 3. The convolution kernel $K(t) := \lambda(t) + 2 \mu(t)$ is a CM function. 

We shall consider the Cauchy problem for equation~\eqref{eq:pressure-attenuation} 
with the initial conditions 
\begin{gather}
P(t,x) = 0 \qquad\text{for $t < 0$}, \label{eq:IC1}\\
\lim_{t\rightarrow 0+} P(t,x) = P_0(x),\quad \lim_{t\rightarrow 0+} P_{,t}(t,x) = Q_0(x)
\label{eq:IC2}
\end{gather}

\section{Dispersion and attenuation theory.}
\label{sec:disp-att}

\subsection{Mathematical preliminaries.}

We shall recall the notion of completely monotonic functions and complete 
Bernstein functions (CBF) and some
properties of the latter class of functions. For more details see 
\cite{BernsteinFunctions} or \cite{HanWM2013}.

\begin{definition}
A real function $f$ defined on $]0,\infty[$ is said to be completely monotonic (CM) if
it is infinitely differentiable and its derivatives satisfy the inequalities
$$(-1)^n \, \D^n \, f(t) \geq 0$$
for $n = 0, 1, 2,\ldots$. 
\end{definition}
A CM function can have a singularity at 0. Any linear combination of CM functions
with positive coefficients is obviously CM. 

If $f$ is CM and it is integrable over $[0,1]$, then $f$ is said to be locally 
integrable CM (LICM). 

According to Bernstein's theorem \cite{GripenbergLondenStaffans} a CM function can 
be expressed in terms of a positive Radon measure:
\begin{theorem} \label{thm:Bernstein}
If $f$ is a CM function then there is a positive Radon measure $m$ such that
\begin{equation} \label{eq:Bernstein}
f(x) = \int_{[0,\infty[} \e^{-x y}\, m(\dd y), \qquad x > 0
\end{equation}
\end{theorem}
For our purposes a Radon measure is essentially a measure with infinite mass.
A CM function $f$ is locally integrable if and only if the measure $m$ satisfies the inequality
$$\int_{[0,\infty[} \frac{m(\dd y)}{1 + y} < \infty$$

A function $f$ on $]0,\infty[$ is said to be a Bernstein function (BF) if it is differentiable 
and its derivative is LICM. A BF is thus non-negative and non-decreasing, hence it has a finite 
limit at 0. 

\begin{definition} 
A real function $f$ on $\overline{\mathbb{R}_+}$ is a CBF if and only if
there is a BF $g$ such that $f(x) = x^2\, \tilde{g}(x)$.
\end{definition}

If $f$ is locally integrable CM, then $p \, \tilde{f}(p) $ is a CBF, where $\tilde{f}$ denotes the 
Laplace transform of $f$. 

The theorems on CBFs cited below can be found in the monograph \cite{BernsteinFunctions}. 

\begin{theorem} \label{thm:xyc}
Every CBF $f$ has an integral representation
\begin{equation} \label{eq:xyc}
f(x) = a + b\, x + x \int_{]0,\infty[} \frac{\nu(\dd r)}{x + r}, \qquad x \geq 0
\end{equation}
with $a, b \geq 0$ and a positive Radon measure $\nu$ satisfying the inequality
\begin{equation} \label{eq:xyb}
\int_{]0,\infty[} \frac{\nu(\dd r)}{1 + r} < \infty
\end{equation}

\end{theorem} 
The integral in equation~\eqref{eq:xyc} is a decreasing function of $x$. Consequently
\begin{equation}
\lim_{x\rightarrow\infty} f(x)/x = b
\end{equation}
while
\begin{equation}
\lim_{x\rightarrow 0} f(x) = a
\end{equation}

\begin{theorem} \label{thm:xf}
A non-zero function $f$ is a CBF if and only if the function 
$x/f(x)$ is a CBF.
\end{theorem}

\begin{theorem} \label{thm:falpha}
If $f$ is a CBF and $0 \leq \alpha \leq 1$ then $f(x)^\alpha$ is
also a CBF.
\end{theorem}

\begin{definition}
A real function $f$ defined on $[0,\infty[$ is slowly varying 
at $w = 0$ or $\infty$ if for all $\lambda > 0$
$$\lim_{x \rightarrow w} f(\lambda\, x)/f(x) = 1.$$
\end{definition}
The logarithm $\ln(1 + x)$ has this property.

\begin{definition}
A real function $f$ defined on $[0,\infty[$ is regularly varying 
at at $w = 0$ or $\infty$ if for all $\lambda > 0$ the limit 
$\lim_{x \rightarrow w} f(\lambda\, x)/f(x) $
is finite.
\end{definition}
If $f$ is regularly varying at $w$ then $f(x) \sim_w x^\alpha\, l(x)$
for some real $\alpha$ and a function $l(x)$ slowly varying at $w$,
where $w$ is either 0 or infinity. 

\begin{theorem}\label{thm:Valiron} (Valiron 1911)\\
If $f$ is an increasing function satisfying the condition 
$\lim_{y\rightarrow -0} f(y) = 0$ and $g(x)$ is given by the Stieltjes integral 
$$g(x) = \int_{[0,\infty[} \frac{\dd f(y)}{x + y}$$
$0 \leq \beta < 1$ and the function $l$ is slowly varying at infinity,  
then the following two statements are equivalent:
\begin{enumerate}[(i)]
\item $f(y) \sim y^\beta\, l(y)$ for $y \rightarrow \infty$;
\item $g(x) \sim [(\uppi \beta)/\sin(\uppi \beta)]\, x^{\beta-1}\, l(x)$.
\end{enumerate}
\end{theorem}
\cite{Shea69}. 

Theorem~\ref{thm:Valiron} can also be applied to integrals
of the form $\int_{[0,\infty[} \mu(\dd y)/(x + y)$ by setting
$f(y) = \mu([0,y])$ for $y \geq 0$ and $f(y) = 0$ if $y < 0$. 

\subsection{Application to the dispersion-attenuation theory.}
\label{sec:appl}

The attenuation-dispersion theory will be presented along the lines of \cite{HanWM2013}.
Although there are some differences between \eqref{eq:pressure-dispersion} below
and eq.~(9) in {\em op. cit.}, the analysis of attenuation and dispersion is very similar. We shall
therefore present the main arguments in brief. Many other results obtained in \cite{HanWM2013}
can be extended to the wave equation under consideration without changing the
argument.

The wavenumber (or the length of the wave number vector) of a pressure field satisfying
equation~\eqref{eq:pressure-attenuation} is given by the formula 
$k = \ii \,\kappa(-\ii \omega)$, where 
\begin{equation} \label{eq:pressure-dispersion}
\kappa(p) = \frac{p}{c_0} \left(1 + p\, \tilde{K}(p)/(\rho_0\, c_0^{\;2})
 \right)^{-1/2}
\end{equation}
The convolution kernel $K(t)$ is assumed to be a locally integrable completely 
monotonic function, hence
$$K(t) = \int_{[0,\infty[} \e^{-r\, t}\, \lambda(\dd r),$$
where $\lambda$ is a positive Radon measure satisfying the
inequality
$$\int_{[0,\infty[} \frac{\lambda(\dd r)}{1 + r} < \infty.$$
Theorem~\ref{thm:xyc} implies that 
$p\, \tilde{K}(p) = p \int_{[0,\infty[} \lambda(\dd r)/(r + p)$ is a CBF. 

By Theorem~\ref{thm:xf} and Theorem~\ref{thm:falpha} the function $\kappa$ is 
a CBF. Consequently the theory of attenuation and dispersion developed for 
linear viscoelastic media applies for the acoustic equation \eqref{eq:pressure-attenuation}. 

Since $\kappa(0) = 0$, Theorem~\ref{thm:xyc} implies that
\begin{equation}\label{eq:linrep}
\kappa(p) = B\, p + \beta(p)
\end{equation}
where $B \geq 0$, 
\begin{equation} \label{eq:beta}
\beta(p) := p \int_{]0,\infty[} \frac{\nu(\dd r)}{r + p}
\end{equation}
and $\nu$ is a positive Radon measure satisfying the inequality  
\begin{equation} \label{eq:ineq}
\int_{]0,\infty[} \frac{\nu(\dd r)}{r + 1} < \infty
\end{equation}
 The support of $\nu$ is called the attenuation spectrum.
Equation~\eqref{eq:beta} implies that
$\beta(p)/p = \mathrm{o}[1]$ for $p \rightarrow \infty$ uniformly in 
the closed right-half complex plane \cite{HanWM2013}. 
This in turn implies that
the wavefield is bounded by a wavefront moving with the speed $C_0 = 1/B$ 
provided $B > 0$. If $B = 0$ then the phase speed is unbounded and the wavefield immediately
spreads to the entire available space. The constant $B$ is given by the limit
$B = \lim_{p\rightarrow \infty} \kappa(p)/p$ (see Appendix A). For some functions
$K(t)$ the Radon measure $\nu$ can be calculated explicitly using the analytic properties of the function $\kappa(p)$ as given by eq.~\eqref{eq:pressure-dispersion}, cf \cite{HanWM2013}.

The function $\beta$ can be split 
into its real part $\mathcal{A}$ 
(attenuation) and imaginary part $-\mathcal{D}$ (excess dispersion). These two 
quantities will be expressed as functions of the circular frequency
$\omega = \ii p$ on the real $\omega$ axis. We then have
\begin{gather}
\mathcal{A}(\omega) = \omega^2 \int_{]0,\infty[\;} 
\frac{\nu(\dd r)}{\omega^2 + r^2} \label{eq:frAtt}\\
\mathcal{D}(\omega) = \omega \int_{]0,\infty[\;} 
\frac{r\, \nu(\dd r)}{\omega^2 + r^2} \label{eq:frDisp}
\end{gather}
Equations~(\ref{eq:frAtt}--\ref{eq:frDisp}) express the two 
functions $\mathcal{A}$ and $\mathcal{D}$ in terms of the same measure $\nu$. They can be 
viewed as a parametric form of the dispersion relations. It will be seen 
that the parametric form of the dispersion relations is often more convenient to use than the 
Kramers-Kronig relations.

It is clear from \eqref{eq:frDisp} that the function $\mathcal{D}$ has sublinear growth 
in the high-frequency range, 
i.~e. $\mathcal{D}(\omega)/\omega \rightarrow 0$ as $\omega \rightarrow \infty$.

Since $y \rightarrow y/(r^2 + y)$ is an increasing function for every $r > 0$,
equation~\eqref{eq:frAtt} implies that the attenuation function $\mathcal{A}(\omega)$ is
increasing unless $\nu = 0$. The statement about the asymptotic behavior of $\beta(p)/p$ at infinity
made previously implies that the attenuation function $\mathcal{A}(\omega)$ is also sublinear 
in the high-frequency range. 

If $\nu$ has a finite total mass
$$M := \int_{]0,\infty[} \nu(\dd r) < \infty$$
then $\lim_{\omega\rightarrow \infty} \mathcal{A}(\omega) = M$ \cite{HanWM2013}. 

The phase speed $c(\omega)$ is defined as $\omega/\re k(\omega)$. 
The definition of the dispersion function $\mathcal{D}$ implies that 
\begin{equation}\label{eq:phase}
\frac{1}{c(\omega)} = \frac{1}{C_0} + \frac{\mathcal{D}(\omega)}{\omega}.
\end{equation}
But $$\frac{\mathcal{D}(\omega)}{\omega} = \int_{]0,\infty[} \frac{r\,\nu(\dd r)}{r^2 + \omega^2}
\geq 0$$
hence $c(\omega) \leq C_0$. If $K_0 := \lim_{t\rightarrow 0} K(t) < \infty$ then $\lim_{p\rightarrow \infty}
[p\, \tilde{K}(p)] = K_0$ in the right half-plane $\re p \geq 0$ and $$C_0 = 
\lim_{p\rightarrow\infty} \frac{p}{\kappa(p)} = c_0 \, \left(1 + K_0/\mathcal{K}\right) \geq c_0.$$
Equation~\eqref{eq:frDisp} implies that $\mathcal{D}(\omega)/\omega$ is a decreasing function of
frequency, hence $c(\omega)$ is an increasing function of $\omega$.
Also $\mathcal{D}(\omega)/\omega \rightarrow 0$ for 
$\omega \rightarrow \infty$,
hence $\lim_{\omega\rightarrow\infty} c(\omega) = C_0$.  

On the other hand
\begin{equation} \label{eq:lim}
\lim_{p\rightarrow 0} \frac{\kappa(p)}{p} = c_0^{\;-1}\, \left( 1 + K_\infty/\mathcal{K}
\right)^{-1/2}
\end{equation}
where $K_\infty := \lim_{t\rightarrow\infty} K(t) = \lim_{p\rightarrow 0} 
[p \, \tilde{K}(p)] \geq 0$. The limit \eqref{eq:lim} exists for arbitrary azimuths $\arg p \in [-\uppi/2,\uppi/2]$,
hence $\lim_{\omega\rightarrow 0} c(\omega) = C_\infty := c_0\, \left(1 + K_\infty/\mathcal{K}\right)^{1/2} > 0$.
We also note that in view of equation~\eqref{eq:phase} for $\omega \rightarrow 0$ the ratio
$\mathcal{D}(\omega)/\omega$ tends to the finite limit $1/C_\infty - 1/C_0$.
Equation~\eqref{eq:frDisp} implies that this limit is equal to $\int_{]0,\infty[} 
\nu(\dd r)/r$. An important conclusion is that the last integral is finite:
\begin{equation} \label{eq:finite}
\int_{]0,\infty[} \frac{\nu(\dd r)}{r} < \infty.
\end{equation}

 Since the function $K$ is non-increasing, the inequality
$K_\infty \leq K_0$ holds and thus $C_\infty \leq C_0$. Hence the phase speed increases 
monotonically from $C_\infty$ to $C_0$.

The parameter $K_\infty$ can be eliminated without changing the model by subtraction
$K_\infty$ from $K(t)$ and adding it to $\mathcal{K}$:
$\mathcal{K} \rightarrow \overline{\mathcal{K}} = \mathcal{K} + K_\infty$,
$K(t) \rightarrow \overline{K}(t) = K(t) - K_\infty$. In terms of $\overline{\mathcal{K}}$ 
and $\overline{K}(t)$ we have $\overline{K}_\infty = 0$, $\overline{C}_\infty = c_0$.

\subsection{Asymptotic behavior of attenuation.}

If $\nu$ has regular variation at $\infty$ 
and $\nu([0,r]) \sim_\infty a \, r^\alpha\, l(r)$, where $l(r)$ is a function slowly varying at $\infty$ 
and $a, \alpha > 0$, then in view of the inequality \eqref{eq:ineq} $\alpha < 1$. It is shown in 
\cite{HanWM2013} that $\mathcal{A}(\omega) \sim  L(\omega)\, \omega^\alpha$,
where $L$ is slowly varying at $\infty$.

If the Radon measure $\nu$ is regularly varying at 0,
$\nu([0,r]) \sim_0 l_1(r)\, r^\gamma$, then in view of the inequality
\eqref{eq:finite} the exponent $\gamma$ must be greater than 1. Theorem~1.11 in 
\cite{HanWM2013} implies that $\mathcal{A}(\omega) \sim_0 \omega^\gamma \,
L_1(\omega)$, where $L_1$ is a slowly varying function at 0. This result is confirmed  
by a frequently reported experimental observation for 
longitudinal waves in polymers and bio-tissues for frequencies in the
range 0--250 \,MHz \cite{SzaboWu00}. This observation is often
misinterpreted as evidence of a power law behavior of attenuation 
$\mathcal{A}(\omega) = A\, \omega^\gamma$, $\gamma > 1$ valid for
all the frequencies.  

In the exceptional case $R : = \int_{]0,\infty[} \nu(\dd r) < \infty$  
the inequality $\nu(\dd r) < \nu(\dd r)/(1 + r^2/\omega^2)$ implies that 
$\lim_{\omega\rightarrow\infty} \mathcal{A}(\omega) = R$. 

Another frequently reported behavior $\mathcal{A}(\omega) \sim a \, \omega^2$ 
for $\omega \rightarrow 0$ (cf \cite{HerzfeldLitovitz,RokhlinAl86})
occurs when the support of the measure $\nu$ is contained in $[b,\infty[$, 
where $b > 0$. In this case
$$\int_{]0,\infty[} \frac{\nu(\dd r)}{r^2 + \omega^2} \rightarrow 
\int_{]b,\infty[} \frac{\nu(\dd r)}{r^2} =: a < \infty$$
as $\omega \rightarrow 0$, which implies the quadratic behavior of attenuation
at low frequencies. According to equation~\eqref{eq:pressure-dispersion}  
this behavior is expected in the case of 
Newtonian viscosity ($K(t) = N\, \delta(t), p \,\tilde{K}(p) = N\, p$) and,
more generally, when $p\, \tilde{K}(p) \sim a \, p$ for $p \rightarrow 0$
and a constant $a$. 

\subsection{Kramers-Kronig dispersion relations.}

Equation~\eqref{eq:linrep} implies that $\beta(p)$ is the
Laplace transform of a causal distribution $f\prime$, where 
$$f(s) = \begin{cases} 
\int_0^\infty \e^{-s r} \, \nu(\dd r), & s > 0\\
\qquad 0, & s \leq 0
\end{cases}
$$ 
Indeed, in view of the inequality \eqref{eq:ineq} the function 
$f$ is locally integrable. Hence its primitive $g$ of $f$ is continuous and the function 
$f$ can be viewed as a distribution of first order.  
Thus $\beta(p) = p \, \tilde{f}(p) = \mathcal{L}(f^\prime)(p)$ is the Laplace 
transform of the causal  distribution $f^\prime$ of second order. If 
$\mathcal{A}(\omega) \sim \mathrm{const} \times \omega^\alpha$ for 
$\omega \rightarrow \infty$, 
where $0 < \alpha < 1$, then 
$\mathcal{A}/(1 + \omega^2)$ and $\mathcal{D}(\omega)/(1 + \omega^2)$ are integrable. 
Since these functions are the real and imaginary part of a Fourier transform
of a causal function, they satisfy the Kramers-Kronig dispersion relations with
one subtraction (\cite{Nussenzveig}, Sec. 1.8(f), pp. 42--43). 

It is however more convenient 
to work with the parametric Kramers-Kronig relations (\ref{eq:frAtt}--\ref{eq:frDisp})
because they do not involve singular integrals. 

\subsection{Regularity of Green's function at the wavefronts.}
\label{sec:regularity}

Low-frequency asymptotics of the attenuation function can be verified by experiments.
High-frequency asymptotics of $\mathcal{A}(\omega)$ is not accessible to such a verification but it determines
the regularity at the wavefronts. The influence of the singularity of the relaxation 
modulus and its derivative at 0 has been examined in several 
papers \cite{Renardy82,HrusaRenardy85,HrusaRenardy86,HrusaRenardy88,DeschGrimmer86,DeschGrimmer89b,HanSerWM,HanQAM}. In this section we shall however link regularity at the wavefront to 
the asymptotic behavior of the attenuation function at infinity.

If $C_0 = \infty$ then Green's function does not vanish anywhere and is an analytic
function of $t$ and $\x$ \cite{Renardy82}. In the remaining cases Green's function
vanishes outside the wavefront $\vert \x \vert = C_0 \, t$ and is an analytic function of $t$ and $\x$ 
in the region of space-time defined by the inequality $t - \vert \x \vert/C_0 > 0$. The 
transition between this region and the unperturbed region is determined
by the discontinuities of Green's function and its derivatives at the wavefront or 
lack thereof. We shall say that the order of the wavefront singularity is $N$ if
if at least one derivative $\partial_t^{\;m} \, \partial_x^{\;n} \, P(t,x)$ 
with $n + m = N$ has a jump discontinuity at the wavefront and this statement is 
not true for any derivatives with $m + n < N$. Green's function of \eqref{eq:pressure-attenuation}
is given by the solution
\eqref{eq:kuk} for $P_0(x) = 0$,  $Q_0(x) = \delta(x)$
\begin{equation} \label{eq:Four}
P(t,x) = \frac{1}{4 \uppi} \int_{-\infty}^\infty
F_1(-\ii \omega)\, \e^{-\ii \omega (t - \vert x \vert/c(\omega)) - \mathcal{A}(\omega) 
\,r} \, \dd \omega
\end{equation}
with $F_1(p) = \kappa(p)/p$ and $r := \vert x \vert$. Note that 
$\vert\kappa(-\ii \omega)/\omega\vert^2 = c(\omega)^{-2} + 
\mathcal{A}(\omega)^2/\omega^2$. The second term is bounded for $\omega > \Omega$, 
where $\Omega$ is an arbitrary positive number,
 because $\mathcal{A}(\omega)$ is sublinear. The first term is bounded by 
$c(\Omega)^{-2}$ because $c(\omega)$ is non-decreasing. 
Let $W_0(\omega,r) := C_1 \, \exp(-\mathcal{A}(\omega)\, r)$. 
The absolute value of $F_1$ is bounded for $\omega > \Omega$ and the absolute value 
of the integrand is majorized by the function $W_0(\omega,r)$, 
for $\omega > \Omega$, where $C_1$ is a positive constant.

Distributional derivatives of $P(t,x)$ are given by the inverse Fourier transform
\begin{equation} \label{eq:derP}
\partial_t^{\;m}\, \partial_x^{\;n} \, P(t,x) = \frac{1}{4 \uppi} \int_{-\infty}^\infty 
F_1(-\ii \omega)
(-\ii \omega)^m\,(-\kappa(-\ii \omega))^n \, \e^{-\ii \omega (t - r/c(\omega)) - 
\mathcal{A}(\omega) \, r} \, \dd \omega
\end{equation}
The absolute value of the integrand is majorized by a function $W_{m,n}(\omega,r) := 
C_2 \, \omega^{m + n} \, \exp(-\mathcal{A}(\omega)\, r)$.

Note that the attenuation function is non-negative and non-decreasing, hence it has a limit 
$\lim_{\omega\rightarrow\infty} \mathcal{A}(\omega) = C$, possibly infinite, $0 \leq C \leq \infty$.
If $C$ is infinite then $\mathcal{A}(\omega)$ can be increasing in the high-frequency range at a logarithmic rate
\begin{equation} \label{eq:log}
\mathcal{A}(\omega) \sim_\infty A\, \ln^{1 + \gamma}(\omega), \qquad A > 0, \gamma > -1
\end{equation}
or according to the power law
\begin{equation} \label{eq:power}
\mathcal{A}(\omega) \sim_\infty A\, \omega^\alpha, \qquad 0 < \alpha < 1
\end{equation}
The attenuation is often bounded in the case of a bounded attenuation spectrum,
which is the case for the Cole-Davidson relaxation \cite{HanWM2013} as well as for the Strick 
and Becker creep \cite{Strick82}. 

We now assume that the wavefront speed $C_0 < \infty$.

\begin{figure}
\begin{center}
\includegraphics[width=0.6\textwidth]{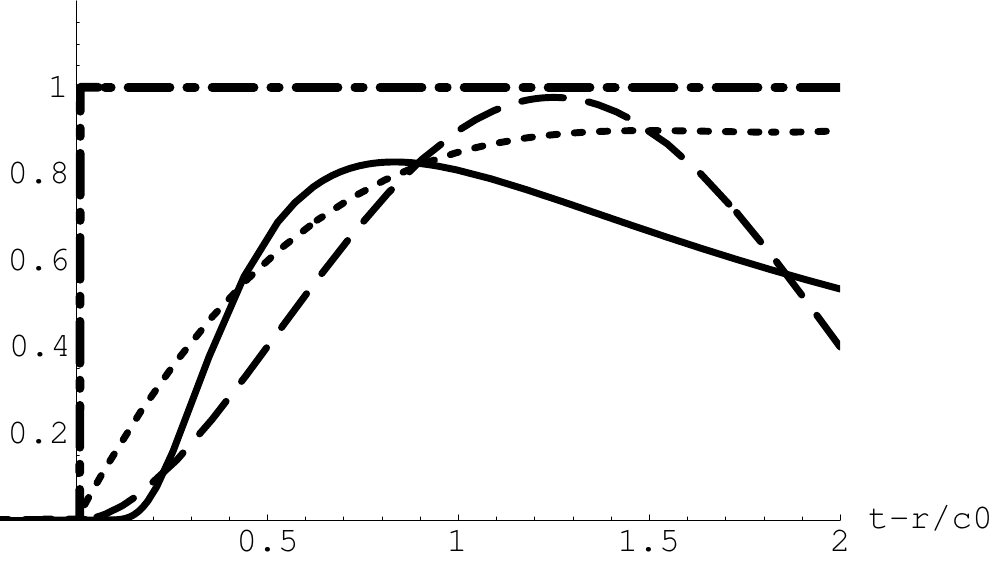}
\end{center}
\caption{Cross-section of Green's function near the wavefront.
(a) Green's function and all its derivatives are continuous at the wavefront (solid line);
(b) Green's function and its first-order derivatives are continuous (dashed line);
(c) Green's function has a jump discontinuity at the wavefront (dot-dashed line);
(d) Green's function is continuous at the wavefront  but its derivatives are not 
(dotted line).} \label{fig:pedestal}
\end{figure}

If the attenuation function satisfies equation~\eqref{eq:power} then for every positive
constant 
$\varepsilon$ and for $r \geq \varepsilon$ the absolute values of the integrands 
of \eqref{eq:Four} and
\eqref{eq:derP} are majorized by the functions
$W_0(\omega,\varepsilon)$ and $W_{m,n}(\omega,\varepsilon)$ are integrable over $[\Omega,\infty[$ and and consequently $P(t,x)$
and its derivatives of arbitrary order are continuous everywhere outside the origin, in particular at the wavefront if $t > 0$. Since they vanish outside
the wavefront, they tend to zero at the wavefront. Consequently the signal propagates 
with a delay with respect to the wavefront and is preceded by a flat "pedestal" 
\cite{Strick1:ConstQ}
(Fig.~\ref{fig:pedestal}). The importance of the pedestal for seismic inversion was demonstrated in \cite{FastTrack}.

If $\mathcal{A}(\omega)$ is bounded then the function $W_0(\omega,r)$ is not integrable and 
Green's function $P(t,x)$ can be discontinuous at the
wavefront. This effect is demonstrated numerically for the Strick and Becker creep in 
\cite{Verweij94}. This effect is frequent if the attenuation spectrum is
bounded, like in the case of relaxation models defined in terms of finite Prony sums,
the Cole-Davidson relaxation \cite{HanWM2013} and the Strick-Becker creep compliance 
\cite{Strick82}.
The appearance of discontinuities at the wavefronts in the last-mentioned case is
demonstrated numerically in \cite{Verweij94}.

A curious intermediary situation arises if $\mathcal{A}(\omega)$ satisfies 
equation~\eqref{eq:log}
with $\gamma = 0$. In this case $W_0(\omega,r) \sim_\infty C_1 \, \omega^{- A\, r}$ is 
integrable if 
the propagation distance is sufficiently large: $r > 1/A$, while
$W_{m,n}(\omega,r) \sim_\infty C_2\, \omega^{m + n - A\, r}$ is integrable if 
$r > (m + n + 1)/A$. At the wavefront $r = c_0\, t$ and thus $P(t,x)$ is continuous
at the wavefront for $t > 1/(A\, c_0)$, while the derivatives of
order $N$ are continuous at the wavefront for $t > (N + 1)/(A \, c_0)$. The order of the 
wavefront singularity thus
increases stepwise with time. This effect was deduced by a different argument by Desch and 
Grimmer \cite{DeschGrimmer86,DeschGrimmer89b}.

If $\gamma > 1$ then $W_0(\omega,r) = C_1 \, \omega^{-A \, r \, \ln^\gamma(\omega)}$ 
and $W_{m,n}(\omega,r) = C_2\, \omega^{m + n - A\, r \,\ln^\gamma(\omega)}$ are integrable for all 
$r \geq \varepsilon > 0$. Consequently $P(t,x)$ and its derivatives are continuous outside
the origin, in particular at the wavefront if $t > 0$. The regularity 
properties of Green's function are thus the same as for power law attenuation.

If $-1 < \gamma < 0$, then $W_0(\omega) = C_1 \, \omega^{-A \, r/\ln^\gamma(\omega)}$ 
and $W_{m,n}(\omega) = C_2\, \omega^{m + n - A\, r/ \ln^\gamma(\omega)}$. In this case both 
$P(t,x)$ and its derivatives can be discontinuous. It follows that strictly logarithmic
growth of attenuation ($\gamma = 1$) constitutes a sharp boundary between media
which allow for discontinuity propagation and those which do not allow for discontinuities
at the wavefronts.

Equation~\eqref{eq:3D} implies that the order of the wavefront in three-dimensional space is
higher by one than in one-dimensional space for the same attenuation function.

\section{Examples.}
\label{sec:examples}

\subsection{Viscoelastic media with a bounded attenuation function.}
\label{sec:bounded}

A special class of viscoelastic relaxation models is characterized by the inequalities 
$K_0 < \infty$ and  
$K_0^\prime := K^\prime(0) > -\infty$. A frequent representative of this class is a
the Dirichlet series
$K(t) = \sum_{n=1}^N \lambda_n \, \e^{-r_n\, t}$, with $N \leq \infty$, 
$\lambda_n, r_n > 0$ for $n = 1,\ldots N$, $\sum_{n=1}^N \lambda_n = K_0< \infty $
and $\sum_{n=1}^N r_n \, \lambda_n = -K_0^\prime < \infty$. A finite Dirichlet series is
known as the Prony series.  
This kind of relaxation mechanism is often used to model multiple
relaxation due to several relaxation mechanisms such as chemical reactions, cf \cite{NachmanAl90}. 
In seismology it was suggested by Liu et al. \cite{LiuKanamoriAnderson}
In this class
$$\int_{]0,\infty[} r\, \lambda(\dd r) = - K_0^\prime < \infty$$
and therefore 
$$p \, \tilde{K}(p) = p \int_{]0,\infty[} \frac{\lambda(\dd r)}{1 + r/p} \approx 
\int_{]0,\infty[} \lambda(\dd r) - \frac{1}{p} \int_{]0,\infty[} r \,\lambda(\dd r) + 
\mathrm{o}[1/p] = K_0 + K_0^\prime/p + \mathrm{o}[1/p]$$
for $p \rightarrow \infty$, $\re p \geq 0$,  so that
$$\frac{\kappa(p)}{p} \approx C_0^{-1} 
+ \frac{R}{p}$$
with $R = -K_0^\prime\, \left[ 1 + K_0/\rho\, c_0^{\;2}\right]^{-3/2} /(2 \rho\, c_0^{\;3})$.
Note that  
$$\frac{\kappa(p)}{p} = \frac{p}{C_0} + \int_{]0,\infty[} \frac{\nu(\dd r )}{1 + r/p} \approx
\frac{p}{C_0} + R$$ for large $p$ implies that
$\int_{]0,\infty[} \nu(\dd r) = R$.
As it has already been noted, this implies that the attenuation function tends to a constant as $\omega \rightarrow \infty$. 

\subsection{Viscoelastic media with a asymptotic power law attenuation and weakly singular $K(t)$.}
\label{sec:powerlaw}

\begin{figure}
\begin{center}
\includegraphics[width=0.7\textwidth]{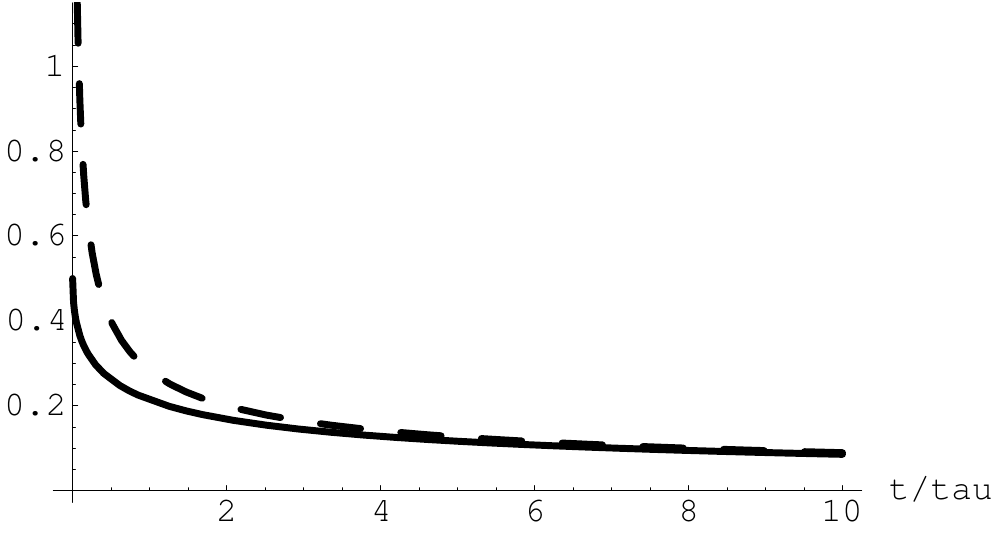}
\end{center}
\caption{Comparison of the kernels $K(t)$ for the Cole-Cole model ($M/\mathcal{K}=1$,
$a = 0.5$) and the constant-$Q$ model ($A/\mathcal{K} = 0.5$).}
\label{plot3}
\end{figure}

The attenuation function grows at a power law rate in the Cole-Cole relaxation model \cite{ColeCole},
originally proposed for dielectric relaxation and subsequently applied in polymer viscoelasticity by  
Bagley and Torvik \cite{BagleyTorvik4}:
\begin{equation}
p\, \tilde{K}(p) = M \, \frac{1 + a \, (\tau\, p)^{-\alpha}}{1 + (\tau\, p)^{-\alpha}} 
- M\, a
\end{equation}
with $M, \tau, a > 0$, $0 < \alpha < 1$. We have subtracted a constant term $M\, a$ so that 
$K_\infty = \lim_{p\rightarrow 0}
[p\, \tilde{K}(p)] = 0$ because a non-zero static modulus is already represented by 
$\mathcal{K}$.  
Since $K_0 = \lim_{p\rightarrow \infty} [p\, \tilde{K}(p)] = M\, (1 - a) \geq 0$, 
the parameter $a$ satisfies the inequality $a \leq 1$. The phase speed is 
contained between $C_\infty > 0$ and  $C_0 < \infty$, where
$C_0:= c_0\, \left[1 + K_0/\mathcal{K} \right]^{1/2} $ is the 
wavefront speed, the phase speed is bounded from below: $c(\omega) \geq c_0$,  
with $c_0 \leq C_0$. The formula
\begin{equation} \label{eq:ML}
\tilde{f}(y) = y^{\alpha-1}/\left(1 + y^\alpha\right)
\end{equation}
 for the Laplace transform of 
$f(x) = \mathrm{E}_\alpha\left(-x^\alpha\right)$ yields the kernel $K$:
\begin{equation} \label{eq:CC}
K(t) = M \,(1 - a) \, \mathrm{E}_\alpha\left(-(t/\tau)^\alpha\right), \qquad t \geq 0, 
\end{equation}
 where $\mathrm{E}_\alpha$ denotes the Mittag-Leffler 
function \cite{PodlubnyBook}. It is proved in Appendix B that $K(t)$ in equation~\eqref{eq:CC}  is CM. This
function is shown in Fig.~\ref{plot3}. The function
has been calculated numerically using equation~\eqref{eq:final}.

The high-frequency behavior of attenuation in the above model is given by the formula 
$$\mathcal{A}(\omega) \sim \frac{(1 - a) \, M_1\, \sin(\alpha \uppi/2)}{2 c_0\, \tau\, 
(1 + M_1)^{3/2}} (\tau \omega)^{1 - \alpha},$$
where $M_1 = M/\mathcal{K}$.
For $\omega \rightarrow 0$ we have a different picture:
$$\mathcal{A}(\omega) \sim \frac{(1 - a) \, M_1\, \sin(\alpha \uppi/2)}{2 c_0\, \tau\, 
(1 + M_1)^{3/2}} (\tau \omega)^{1 + \alpha}$$
The exponent $1 + \alpha$ lies between $1$ and $2$, in accordance with experimental data
for polymers and bio-tissues in the frequency range 0--250\, MHz. 

\begin{figure}
\begin{minipage}{0.45\textwidth}
\includegraphics[width=\textwidth]{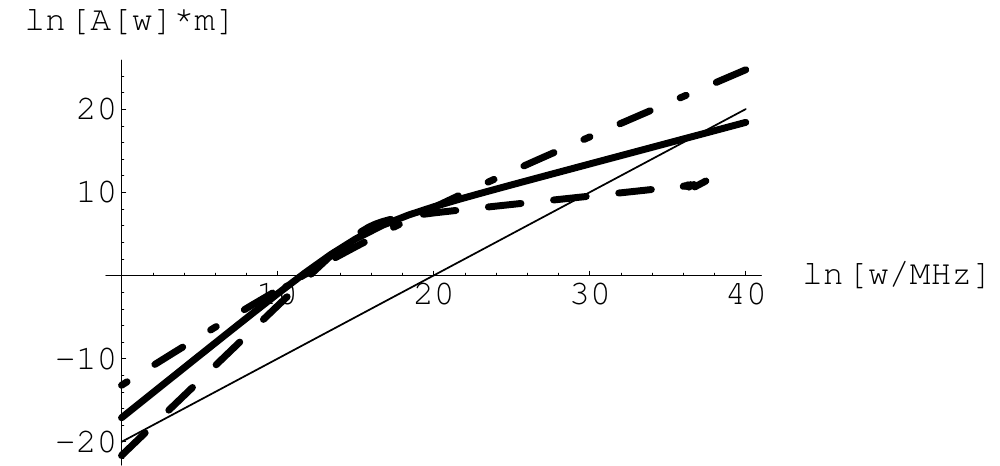}
\begin{center} {\small (a) Log-log plot of the attenuation function.}
\end{center}
\end{minipage}
\begin{minipage}{0.45\textwidth}
\includegraphics[width=\textwidth]{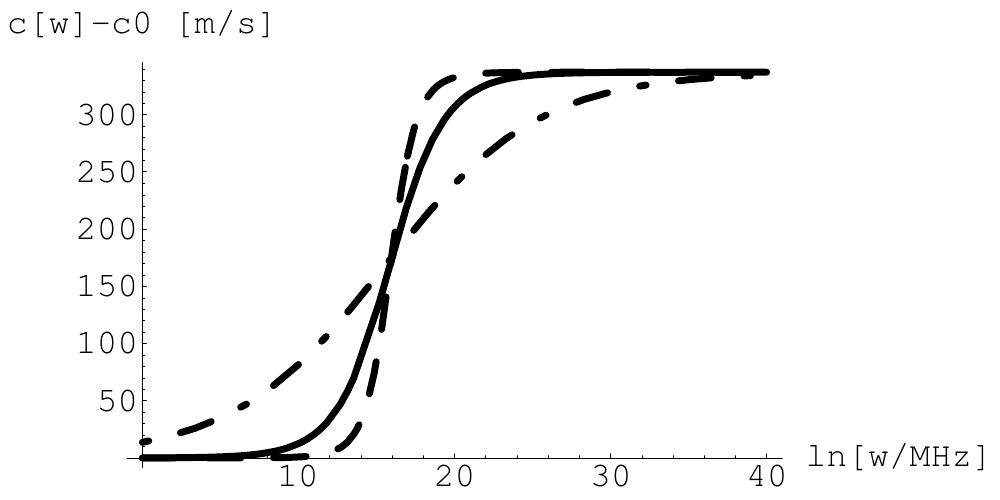}
\begin{center} {\small (b) Plot of phase speed vs $\ln(\omega/\mathrm{MHz})$.}
\end{center}
\end{minipage}
\caption{Attenuation function and phase speed for the Cole-Cole relaxation model for $\alpha =$ 0.2 (dot-dash), 0.5 (solid), 0.8 (dash).} \label{plot1}
\end{figure}

The Cole-Cole attenuation function can be calculated numerically using the formula
\begin{gather}
\mathcal{A}(\omega) = \omega \, \sqrt{\sqrt{X(\omega)^2 + Y(\omega)^2} - X(\omega)}/\left(2 c_0\, \sqrt{X(\omega)^2 + Y(\omega)^2}\right)\\
X(\omega) = 1 + (M*(1 - a)/\mathcal{K}) (\omega\,\tau)^\alpha\, \left((\omega\, \tau)^\alpha + 
\cos(\uppi \alpha/2)\right)/\left(1 + (\omega\, \tau)^{2 \alpha} + 
2 (\omega \, \tau^\alpha \,\cos(\uppi \alpha/2)\right)\\
Y(\omega) = -(M/\mathcal{K})\, (1 - a) \, (\omega \tau)^\alpha \, \sin(\uppi \alpha/2)/
\left(1 + (\omega\, \tau)^{2 \alpha} + 
2 (\omega \, \tau)^\alpha \,\cos(\uppi \alpha/2)\right)
\end{gather}
Using these formulas the functions $\mathcal{A}(\omega)$ and $c(\omega)$ were plotted in Fig.~\ref{plot1}
for $c_0 = 1500 \,\mathrm{m}/\mathrm{s}$, $a = 0.5$, $M/\mathcal{K} = 1$, $K_\infty/\mathcal{K} 
= a$,  
$\tau = 10^{-13} \, \mathrm{s}$ and $\alpha = 0.2, 0.5$ and 0.8.
The plot shows that the Cole-Cole attenuation function obeys two approximate power laws, 
a superlinear one in the low frequency range $\omega \ll 3.27\times 10^6\, \mathrm{MHz}$ and a sublinear
one in the high frequency range $\omega \gg 3.27\times 10^6\, \mathrm{MHz}$.

\subsection{A viscoelastic model with unbounded phase speed.}
\label{sec:stronglysingular}

\begin{figure}
\begin{minipage}{0.45\textwidth}
\includegraphics[width=\textwidth]{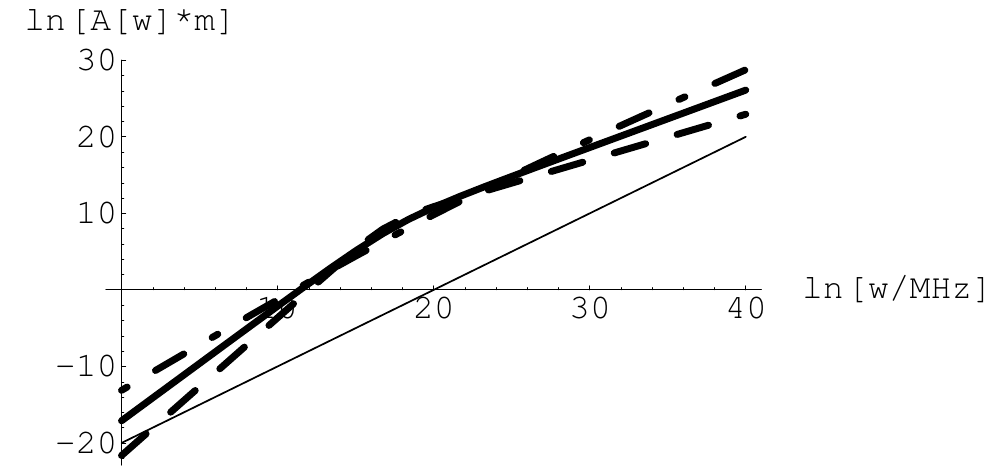}
\begin{center} {\small (a) Log-log plot of the attenuation function.}
\end{center}
\end{minipage}
\begin{minipage}{0.45\textwidth}
\includegraphics[width=\textwidth]{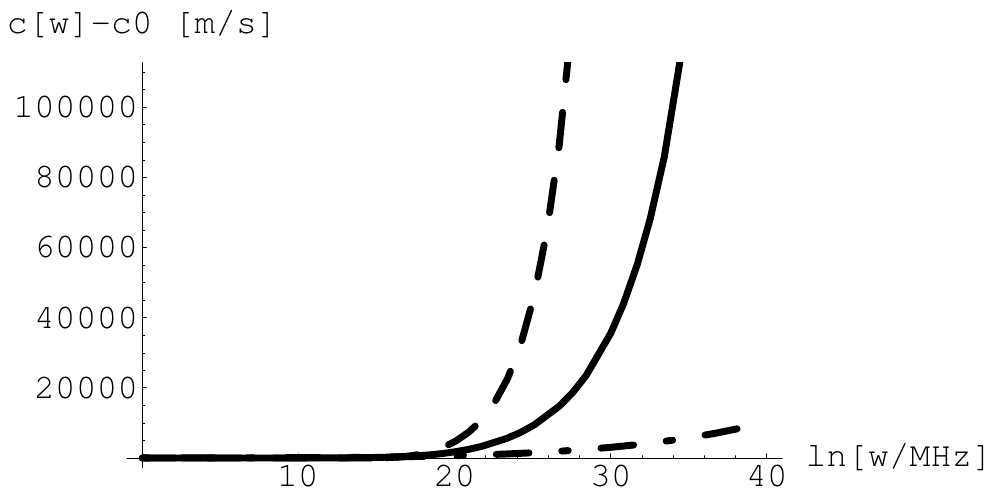}
\begin{center} {\small (b) Plot of phase speed vs $\ln(\omega/\mathrm{MHz})$.}
\end{center}
\end{minipage}
\caption{Attenuation function and phase speed for the constant-Q model for $\alpha =$ 0.2 (dot-dash), 0.5 (solid), 0.8 (dash).} \label{plot2}
\end{figure}

If $K(t) = A \, (t/\tau)^{-\alpha}/\Gamma(1-\alpha)$ with $A, \tau > 0$ and  
$0 < \alpha < 1$, then the pressure equation assumes the form
\begin{equation} \label{eq:constQ}
c_0^{\;-2} \, P_{,tt} = \nabla^2 \, P + (A \, \tau^\alpha/\rho_0\, c_0^{\;2}) \D^\alpha \, \nabla^2 \,P
\end{equation}
where $\D^\alpha$ denotes the Caputo fractional time derivative of order $\alpha$. 
This particular equation was considered in \cite{HolmSinkus2010}.
It is related to the constant-Q model in seismology \cite{Kjartansson:ConstQ,CaCaMaHa}.
In this case $K_0 = \infty$, hence $C_0 = \infty$ and the solutions do not
exhibit wavefronts.  In this case $X(\omega) = 1 + A_1 \, (\tau \omega)^\alpha \, 
\cos(\uppi \, \alpha/2)$ and $Y(\omega) = A_1\, \, (\tau \omega)^\alpha \, 
\sin(\uppi \, \alpha/2)$, where $A_1 := A/\mathcal{K}$. The attenuation function 
and phase speed are shown for $A_1 = 0.5$ in Fig.~\ref{plot2},
which confirms that the phase speed is unbounded. This is hardly surprising because the
order of the derivatives in the last term of equation~\eqref{eq:constQ} is 
$2 + \alpha$, higher than the orders of the other derivatives and the equation of motion is parabolic. 

Asymptotic behavior of dispersion and attenuation associated with special wave 
equations based on fractional versions of spring-dashpot models 
has been examined in a series of papers by Holm, N\"{a}sholm and 
Sinkus \cite{HolmSinkus2010,NasholmHolm2011,NasholmHolm2012,NasholmHolm2013}. The
convolution kernels $K$ corresponding to the fractional spring-dashpot models are CM, hence 
the general theory developed in Sec.~\ref{sec:appl} applies to their equations.

\subsection{Quasi-linear attenuation.}
\label{sec:ql}

The attenuation function is almost linear while phase speed varies very slowly  
in seismological applications \cite{Futterman,Strick2:constQ} as well as in marine 
sediments \cite{Buckingham2007}. This results in an approximately constant $Q$ factor, 
defined by the formula $Q(\omega) := \omega/[2 c(\omega)\, \mathcal{A}(\omega)]$. 

We shall use asymptotic considerations to investigate dispersion,
attenuation and existence of wavefronts for a nearly linear attenuation function. 
Note that an exactly linear rate of growth of the attenuation function
in the high frequency range would be inconsistent with the
assumption that the origin of attenuation is purely viscoelastic and
the relaxation modulus is CM. A model of a linear attenuation function
and an approximately linear $\mathcal{D}(\omega)$ was elaborated by
Futterman \cite{Futterman}. 
Futterman was only concerned with finite phase speed for his strictly linear 
attenuation model. In contrast to Futterman we shall take into account the fact that 
attenuation grows at a strictly sublinear rate. Furthermore, 
the investigations of this section also shed some light on the relation between 
boundedness of
the phase speed and existence of wavefronts. One might wonder whether it is possible that
the phase speed has a finite upper bound $C_0$ but the wave field extends 
beyond the surface $\vert\mathbf{x}\vert = C_0 \, t$. It will turn out that this can
happen. 

Note that the 
attenuation function in the constant-$Q$ model of the previous section 
is sublinear, while absence of wavefronts manifests itself through unboundedness 
of phase speed (Fig.~\ref{plot2}). Sublinearity of the attenuation function alone 
does not however guarantee 
bounded phase speed. The Fourier transform of the one-dimensional Green's function 
is $$g(\omega) = F(\omega) \, \e^{-\ii \, \omega\,(t - \vert \x\vert/C_0) - \ii\, \mathcal{D}(\omega)\, \vert \x \vert
-\mathcal{A}(\omega)\, \vert \x \vert}$$
where $\vert F(\omega)\vert$ is bounded. 
The function $g$ is square integrable for $\vert \x \vert > 0$ if
$\mathcal{A}(\omega)/\vert\ln(\omega)\vert$ is unbounded for $\omega \rightarrow \infty$.
Suppose that this condition is satisfied. 
According to the Paley-Wiener theorem  (\cite{PaleyWiener}, Theorem~XII) Green's 
function vanishes for $t < \vert \x \vert/C_0$ if and only if 
$$ \int_0^\infty \frac{\mathcal{A}(\omega)}{1 + \omega^2}\, \dd \omega < \infty$$
This is true in particular for 
\begin{equation} \label{eq:as}
\mathcal{A}(\omega) \sim a\, \omega^{1 + \lambda}/\vert 
\ln(\omega)\vert^\gamma, \qquad a > 0, \quad\omega \rightarrow \infty \end{equation}
 if the following condition is satisfied:\\ \vspace{0.2cm}
$(\ast)$ Either $\lambda < 0$ or $\lambda = 0$ and $\gamma > 1 + \varepsilon$. 

In order to check the 
behavior of attenuation and phase speed in these cases 
we have to turn to equations~\eqref{eq:frAtt} and \eqref{eq:frDisp}.  
Consider a Radon measure $\nu(\dd r) = h(r) \, \dd r$, whose density has the 
asymptotic behavior $h(r) \sim b\, r^\lambda/\vert \ln r\vert^\gamma$ at infinity,
where $b > 0$.
Inequality~\eqref{eq:ineq} is satisfied if
$$b \int_N^\infty \frac{r^\lambda\, \dd r}{(1 + r)\, \ln(r)^\gamma} \equiv
b \int_{\e^N}^\infty \frac{\e^{\lambda \, y} \, \dd y}{y^\gamma \, ( 1 + \e^{-y})} < \infty$$
for some sufficiently large $N > 1$, hence it is satisfied if and only if
Condition~($\ast$) is satisfied. 

We shall now calculate the 
asymptotic behavior of the functions $\mathcal{A}(\omega)$ and $\mathcal{D}(\omega)$.
In terms of the integration variable $s = r^2$
$$\mathcal{A}(\omega) = \omega^2 \int_{]0,\infty[} \frac{\mu(\dd s)}{\omega^2 + s}$$
where $\mu(\dd s) = h\left(\sqrt{s}\right)\, \dd s/(2 \sqrt{s})
\sim b\, s^{(\lambda-1)/2}\, \dd s/\left(2 \vert \ln(\sqrt{s})\vert^\gamma\right)$
for $s \rightarrow \infty$. Note that $1/\vert \ln(\sqrt{s})\vert^\gamma$ is a slowly varying
function. 
By Theorem~\ref{thm:Valiron} $$\mathcal{A}(\omega) \sim (b/2)\, [\uppi/\cos(\uppi \lambda/2)]\,\omega^{1+ \lambda}/
\vert \ln(\omega)\vert^\gamma$$ for $\omega \rightarrow \infty.$ 
It is thus seen that in the case under consideration 
inequality \eqref{eq:ineq} is equivalent to the statement that $\vert \x \vert = C_0\, t$ 
is a wavefront of Green's function and also to Condition~($\ast$). 

By a similar argument equation~\eqref{eq:frDisp} 
implies that 
$$\mathcal{D}(\omega) \sim (b/2) [\uppi/\cos(\uppi \lambda/2)]\,\omega^{1 + \lambda}/\vert \ln(\omega) \vert^\gamma$$ 
and, using equation~\eqref{eq:phase},
$$c(\omega) \sim C_0 \, \left[1 - b \omega^\lambda/(2 \vert \ln(\omega)\vert^\gamma\right]$$
If $\lambda < 0$ or $\lambda = 0$ and $\gamma > 0$, then 
$\lim_{\omega\rightarrow\infty} c(\omega) = C_0$. 

If $\lambda = 0$ and $0 < \gamma \leq 1$ then the phase speed has a finite upper
bound $C_0$ but the Paley-Wiener theorem implies that 
the wavefield is not bounded by the surface $\vert \x \vert = C_0\, t$. It should however
be noted that in this special case failure of inequality~\eqref{eq:ineq} implies
that the kernel $K(t)$ cannot be locally integrable and CM. For $\lambda = 0$ and 
$\gamma > 0$ the dispersion is 
very weak. A nearly linear attenuation and a nearly constant phase speed is observed in marine sediments as well as in the Earth's crust and mantle. 

\section{Conclusions.}

Consistency with linear viscoelasticity requires that attenuation in the pressure 
wave equation
is represented by a term of the form $K\ast \nabla^2 \, P$. The operator $\nabla^2$ is
here crucial for truly viscoelastic attenuation. 

If the kernel $K$ is completely monotonic then the wavenumber function $\kappa(p)$
is a CBF and a rich theory of dispersion-attenuation developed in \cite{SerHanJMP,HanWM2013}
applies. The general results presented in Sec.~\ref{sec:appl} has a very strong predictive 
power even before any particular model of the medium is substituted.

It was pointed out in Section~\ref{sec:appl} that $\mathcal{A}(\omega)$ is sublinear 
in the high-frequency range. This is also true when phase speed is unbounded and 
there are no wavefronts.  As
the results of Sections~\ref{sec:appl} and \ref{sec:ql} indicate, absence of 
wavefronts manifests itself
by unboundedness of the phase speed rather than by the Paley-Wiener criterion alone.
On the other hand experimental data for many polymers, castor oil and bio-tissues 
indicate a power law 
for the attenuation function $\mathcal{A}(\omega)$ with an exponent in the range 1--2.
Such values of the exponent in the high frequency range are incompatible with
viscoelasticity but they 
are consistent with the low frequency asymptotics of the attenuation function. They are also 
incompatible with bounded phase speed. Viscoelasticity does not however 
exclude an unbounded phase speed.   
 Comparison of characteristic 
relaxation times with the range of experimental data obtained for polymers and 
bio-tissues (0--250 \, MHz) indicate that 
the observed data are pertinent for the low-frequency behavior of the attenuation function.
Low-frequency behavior of attenuation and dispersion is examined in some more detail 
in the context of linear viscoelasticity in \cite{HanWM2013}. 

High-frequency asymptotic properties of the attenuation function are unavailable to direct
measurements but they affect the singularity carried by the wavefront if $C_0 < \infty$.

\appendix

\section{Solution of the Cauchy problem~(\ref{eq:pressure-attenuation}--
\ref{eq:IC2}}

Consider the initial value problem defined by equation~\eqref{eq:pressure-attenuation} 
in one-dimensional space with the initial conditions $P(0,x) = P_0(x)$ and 
$P_{,t}(0,x) = Q_0(x)$. The Laplace-Fourier transform 
$$\hat{\tilde{P}}(p,k) := \int_0^\infty \e^{-p t} \left[ \int_{-\infty}^\infty 
\e^{-\ii k x} \, P(t,x) \, \dd x\right] \, \dd t$$
is given by the expression
$\hat{\tilde{P}}(p,k) = F(p,k)/\left[k^2 + \kappa(p)^2\right]$, 
where 
$$F(p,k) := c_0^{\;-2}\,\frac{\left[p\, (1 + k^2/\rho)\, \hat{P}_0(k) + \hat{Q}_0(k)\right]}{1 + 
\left(\rho_0\, c_0^{\;2}\right)^{-1} \, p \, \tilde{K}(p)} 
$$
Hence
$$P(t,x) = \frac{1}{2 \uppi \ii} \int_{-\ii \infty+\varepsilon}^{\ii \infty+\varepsilon}
\e^{p t}\, \left[\frac{1}{2 \uppi} \int_{-\infty}^\infty \e^{\ii k x} F(p,k)
\, \frac{1}{k^2 + \kappa(p)^2} \, \dd k\right] \, \dd p$$
for an arbitrary $\varepsilon > 0$.

If $x > 0$/$x < 0$ then the contour of the inner integral can be closed by a semicircle 
at infinity in the upper/lower complex $k$-half-plane, $\im k > 0$/$\im k < 0$.
Since $\re \kappa(-\ii \omega) = \mathcal{A}(\omega) \geq 0$, the only residue
in the upper/lower $k$-half-plane is $k = \ii \kappa(-\ii \omega)$/
$k = -\ii \kappa(-\ii \omega)$,
hence 
\begin{equation} \label{eq:kuk}
P(t,x) = \frac{1}{4 \uppi \ii} \int_{-\ii \infty+\varepsilon}^{\ii \infty+\varepsilon}
F_1(p)\, \e^{p t - \kappa(p) \, \vert x \vert} \, \dd p = 
\frac{1}{4 \uppi \ii} \int_{-\ii \infty+\varepsilon}^{\ii \infty+\varepsilon}
F_1(p)\, \e^{p (t - B \,\vert x \vert) - \beta(p) \vert x \vert} \, \dd p
\end{equation}
where $F_1(p) := F(p,\kappa(p)\, \sgn x)/\kappa(p)$.

For $t < B \, \vert x \vert$ the Bromwich contour can be closed in the right $p$-half-plane
$\re p > 0$. The function $\kappa(p)$ does not have any singularity except for a cut
along the negative $p$-semi-axis. 
The integrand does not have any singularity in the right-half plane, hence  
$P(t,x) = 0$ for $t < B \, \vert x \vert$. If $B > 0$ then $C_0 := 1/B$ can be
identified as the wavefront speed. If $B = 0$ then $F(t,x)$ vanishes for $t < 0$,
which implies causality.

Define the function $Q(t,y)$ in such a way that $P(t,x) = Q(t,\vert x \vert)$. 
The solution $P^{(3)}(t,\x)$ of the same initial-value problem in three dimensions is then
given by the formula \cite{TimeFrac} 
\begin{equation} \label{eq:3D}
P^{(3)}(t,\x) = - \frac{1}{2 \uppi r} 
\frac{\partial Q(t,r)}{\partial r}
\end{equation} 
where $r = \vert \x \vert$, and therefore $P^{(3)}(t,\x)$ vanishes for 
$t < B\, r$.

\section{The Cole-Cole relaxation kernel.}

The Mittag-Leffler function $\mathrm{E}_\alpha\left(-x^\alpha\right)$,
$0 < \alpha \leq 1$, can be calculated 
by applying the inverse Laplace transform to equation~\eqref{eq:ML}:
$$\mathrm{E}_\alpha\left(-x^\alpha\right) = \frac{1}{2 \uppi \ii}
\int_{\varepsilon - \ii \infty}^{\varepsilon + \ii \infty} 
\e^{y\, x} \frac{y^{\alpha-1}}{y^\alpha + 1} \dd y$$
where $\varepsilon > 0$. The Bromwich contour can be deformed to a Hankel loop
running along a the straight line $\mathcal{L}_-$  from $-\infty-\ii \varepsilon$ to $-\ii \varepsilon$ , then along a semicircle 
of radius $\varepsilon$ centered at 0 and along the straight line $\mathcal{L}_+$ from $\ii \varepsilon$ to $-\infty +
\ii \varepsilon$. The integral over the semi-circle 
$$\int_{-\uppi}^\uppi \e^{\varepsilon x \, \exp(\ii \varphi)}\,
\frac{\varepsilon^{\alpha-1} \exp(\ii \,(\alpha-1)\, \varphi)}{1 + 
\varepsilon^\alpha\, \exp(\ii \,\alpha\, \varphi)} \ii \varepsilon \, \e^{\ii \varphi}\,
\dd \varphi$$
tends to 0 as $\varepsilon \rightarrow 0$.
The remaining two integrals tend to
\begin{equation} \label{eq:xxx}
\frac{1}{\uppi} \int_0^\infty \e^{-r t} \,\im\left[ \frac{r^{\alpha-1} \, 
\exp(-\ii\, (\alpha - 1)\, \uppi)}{1 + r^\alpha \, 
\exp(-\ii\, \alpha \, \uppi)}\right] \, \dd r
\end{equation}
where $r \, \exp(\pm \ii \uppi)$ has been substituted for $y$ on 
$\mathcal{L}_\pm$. 
Equation~\eqref{eq:xxx} works out to
\begin{equation} \label{eq:final}
\mathrm{E}_\alpha\left(-x^\alpha\right) = 
\frac{\sin(\alpha\, \uppi)}{\uppi} \int_0^\infty 
\e^{-r\, x} \, r^{\alpha - 1} \, \left[ r^{2\, \alpha} + 2 \,r^\alpha\, \cos(\alpha\, \uppi) + 1\right]^{-1}\, \dd r
\end{equation} 

Equation~\eqref{eq:xxx} shows that $\mathrm{E}_\alpha\left(-x^\alpha\right)$ is
the Laplace transform of a non-negative function. By Theorem~\ref{thm:Bernstein} 
it is CM.

The Cole-Cole kernel function \eqref{eq:CC} differs from 
$\mathrm{E}_\alpha\left(-x^\alpha\right)$
by a linear scaling transformation, hence it is CM.

\end{document}